\RequirePackage[2020-02-02]{latexrelease}
\documentclass[twocolumn,preprintnumbers]{revtex4}%
\usepackage{amssymb}
\usepackage{amsmath}
\usepackage{graphicx}
\usepackage{dcolumn}
\usepackage{bm}
\usepackage{xcolor}
\usepackage{amsfonts}%
\UseRawInputEncoding
\RequirePackage[2020-02-02]{latexrelease}
\begin{document}
\title{Frequency mixing spectroscopy of spins in diamond}
\author{Mohammed Attrash}
\affiliation{Andrew and Erna Viterbi Department of Electrical Engineering, Technion, Haifa
32000, Israel}
\author{Sergei Masis}
\affiliation{Andrew and Erna Viterbi Department of Electrical Engineering, Technion, Haifa
32000, Israel}
\author{Sergey Hazanov}
\affiliation{Andrew and Erna Viterbi Department of Electrical Engineering, Technion, Haifa
32000, Israel}
\author{Oleg Shtempluck}
\affiliation{Andrew and Erna Viterbi Department of Electrical Engineering, Technion, Haifa
32000, Israel}
\author{Eyal Buks}
\affiliation{Andrew and Erna Viterbi Department of Electrical Engineering, Technion, Haifa
32000, Israel}
\date{\today }

\begin{abstract}
Frequency mixing processes in spin systems have a variety of applications in
meteorology and in quantum data processing. Spin spectroscopy based on
frequency mixing offers some advantages, including the ability to eliminate
crosstalk between driving and detection. We experimentally explore nonlinear
frequency mixing processes with negatively charged nitrogen-vacancy defects in
diamond at low temperatures, and near level anti crossing. The experimental
setup allows simultaneously applying magnetic driving in the longitudinal and
transverse directions. Magnetic resonance detection is demonstrated using both
Landau Zener St\"uckelberg interferometry and two-tone driving spectroscopy.
The experimental results are compared with predictions of a theoretical
analysis based on the rotating wave approximation.

\end{abstract}
\maketitle

\textbf{Introduction} - Spins are essentially the most nonlinear systems found
in nature. Their magnetic resonance is widely employed for sensing and imaging
applications. In some cases, nonlinear response impacts the sensitivity and
bandwidth of magnetic resonance imaging systems
\cite{Yamaguchi_100901,Alfasi_063808}. Nonlinear response can be exploited for
performance enhancement in some cases. One example is enhancement of magnetic
resonance detection sensitivity that is achieved using squeezed microwave
fields \cite{Bienfait_041011}. Another example is driving-induced enhancement
of coupling between spins \cite{Levi_053516}. Moreover, for quantum data
processing, nonlinear response is an essential resource enabling the
generation of entangled states \cite{Gottesman_012310} and topological
frequency conversion \cite{Martin_041008}. In addition, nonlinear response of
driven spins can be exploited for the generation of exotic states of matter by
means of Floquet engineering \cite{Eckardt_093039}.

Frequency mixing processes that are enabled by nonlinear response can be
employed for a variety of applications. In spectroscopy, sensitivity is
commonly limited by crosstalk, when both driving and detection are performed
at the same frequency (which is usually the case in cavity-based
spectroscopy \cite{Alfasi_063808}). This problem can be avoided
by employing frequency mixing, which enables resonance driving using only
off-resonance driving tones. Other examples for frequency mixing applications
are electromagnetically induced transparency \cite{Manson_1659} and hole
burning \cite{Kehayias_245202}. Frequency mixing can be used to manipulate
transition rates between quantum states, which, in turn, can enable
controlling states' population. Cooling induced by frequency mixing has been
demonstrated in \cite{Oliver_261} using a Josephson flux qubit. A similar
process of frequency mixing can be implemented to achieve population inversion.

Here we explore frequency mixing processes in negatively charged
nitrogen-vacancy (NV$^{-}$) defects in diamond at low temperatures. A NV$^{-}$
defect has a spin triplet ground state with magnetic quantum numbers
$m_{\mathrm{S}}\in\left\{  -1,0,1\right\}  $ \cite{Doherty_1}. The NV$^{-}$
electronic spin state can be polarized and read out with light in the optical
band. Dense ensembles of NV$^{-}$ defects were demonstrated to be applicable
for magnetometry~\cite{Rondin_056503}, classical~\cite{Dhomkar_e1600911} and
quantum \cite{Zhu2011,Kubo2011,Amsuss2011,Grezes_021049} information storage,
and for a maser implementation \cite{Breeze_493}. The NV magnetometry
sensitivity may reach sub picotesla per $\operatorname{Hz}^{1/2}$ level
\cite{Wolf_041001,Alsid_054095,Wang_021061}. Dual driving of NV$^{-}$
electrons and nuclear spins were investigated by optical detection of magnetic
resonance (ODMR)
\cite{Dmitriev_011801,Dmitriev_1,Dmitriev_043509,Childress_033839,Clevenson_021401,Mamin_030803,Wang_021061}%
, Landau Zener St\"uckelberg interferometry \cite{Zhou_010503,Huang_011003},
and electron-spin double resonance \cite{Yamaguchi_100901,Mikawa_012610}.

In this work we explore mixing between longitudinal driving in the radio
frequency (RF) band and transverse driving in the microwave (MW) band. To
account for our experimental results, theoretical predictions derived using
the rotating wave approximation (RWA) are presented. We find relatively strong
nonlinear response near the NV$^{-}$ ground state level anti crossing (GSLAC)
\cite{Jacques_057403,Childress_033839,Kehayias_245202,Fuchs_789,Clevenson_021401}%
. The response enhancement is attributed to a significant state mixing
occurring near GSLAC. The same state mixing gives rise to a significant
modification in transition selection rules. In particular, the commonly
forbidden transition between $m_{\mathrm{S}}=-1$ and $m_{\mathrm{S}}=+1$
states, becomes partially allowed near GSLAC. This transition can be used to
double spin detection sensitivity (compared to sensitivity obtained using
transitions associated with a change in the magnetic quantum number
$m_{\mathrm{S}}$ equals $+1$ or $-1$). However, away from GSLAC, detection
based on this transition requires double-quantum driving \cite{Mamin_030803}.

\textbf{Experimental setup} - A diamond sample with nitrogen concentration
lower than 200 ppm was electron irradiated at energy and dose of 2.8 MeV and
$8\times10^{18}%
\operatorname{cm}%
^{-2}$, respectively, and annealed at $900^{\circ}$C for 2 hours to create
NV$^{-}$ defects. The NV$^{-}$ concentration is $3\times10^{17}%
\operatorname{cm}%
^{-3}$. Concentration was estimated by comparing with a reference sample
having a known NV concentration (with sufficiently low laser power, for which
fluorescence is proportional to NV concentration) \cite{Farfurnik_123101}. The
sample assembly was placed at a cryostat with a base temperature of $3.6%
\operatorname{K}%
$. To reach the GSLAC region, the diamond [111] axis direction was placed
nearly parallel to the externally applied static magnetic field, which was
generated by superconducting coils.

\begin{figure}[ptb]
\begin{center}
\includegraphics[width=3.2in,keepaspectratio]{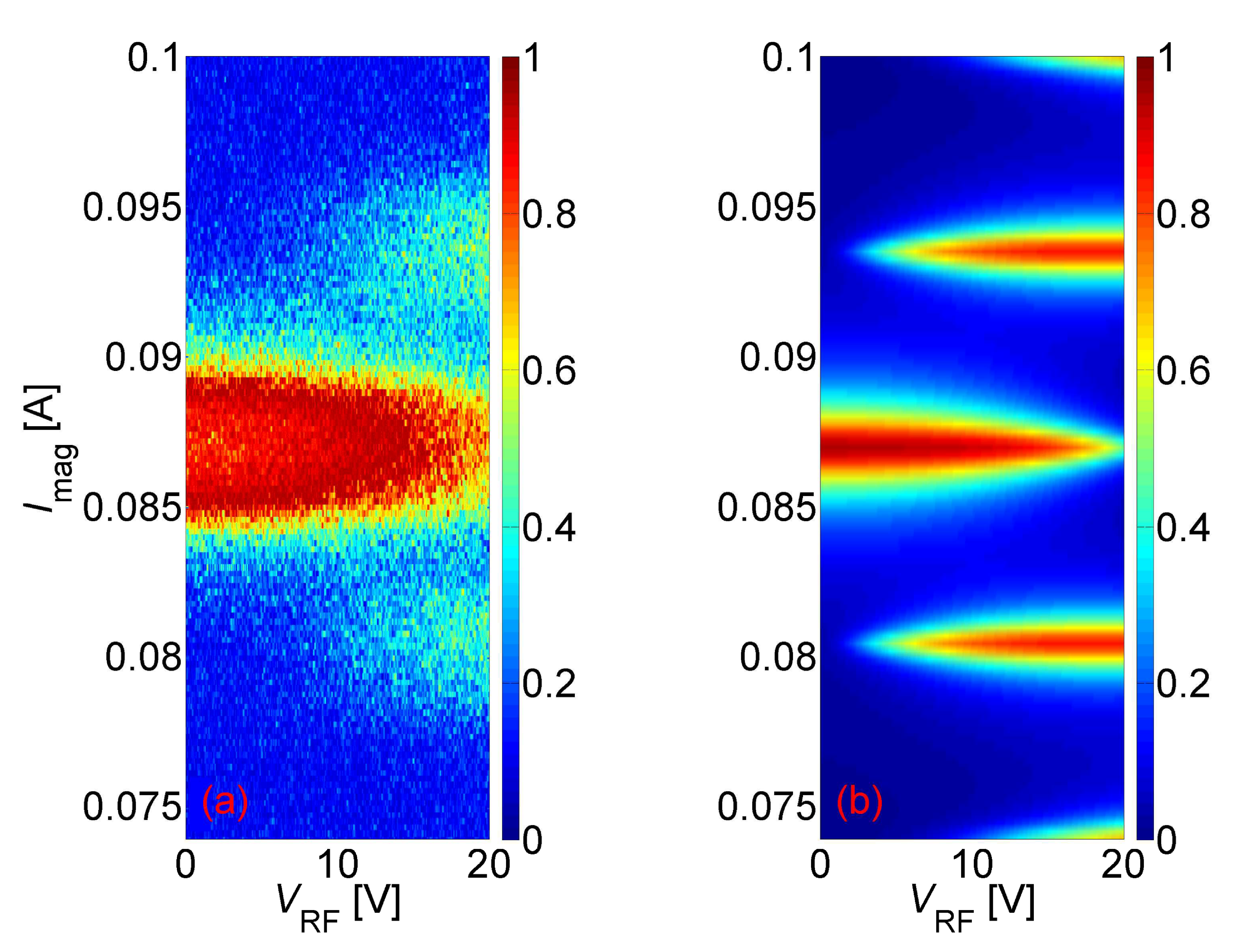}
\end{center}
\caption{Two antenna drive. (a) ODMR signal as a function of RF solenoid
applied voltage amplitude $V_{\mathrm{RF}}$, and current applied to the
solenoid generating the static magnetic field $I_{\mathrm{mag} }$. The fixed
RF solenoid driving frequency is $\Omega_{\mathrm{L}}/\left(  2\pi\right)
=10.5 \operatorname{MHz} $, MW LA driving amplitude is $\Omega_{\mathrm{T}
1}=5.6 \operatorname{MHz}$, and MW LA driving frequency is $\Omega
_{\mathrm{T}}/\left(  2\pi\right)  =3.15 \operatorname{GHz} $. (b) Calculated
driving-induced polarization coefficient $\mathcal{P}$ [see Eq. (\ref{P_n1,n2}%
)]. The parameters' assumed values for the calculation of $\mathcal{P}$ are
$\gamma_{1}=0.5\operatorname{MHz}$ and $\gamma_{2}=2\operatorname{MHz}$.}%
\label{FigLZd}%
\end{figure}

Two antennas are employed for spin driving. An RF solenoid, having axis nearly
parallel to the applied static magnetic field, allows longitudinal driving,
whereas transverse driving is applied using a microwave loop antenna (LA)
having axis nearly orthogonal to the applied static magnetic field. The
impedance mismatching coefficient $\zeta$, which is given by $\zeta=$ $\omega
L/Z_{0}$ \cite{Pozar_MWE}, where $\omega$ is the driving angular frequency,
$L$ is the antenna's inductance, and $Z_{0}=50%
\operatorname{\Omega }%
$ is the impedance of the coaxial cable attached to the antenna, is $\zeta=94$
for the RF solenoid at $\omega/2\pi=10%
\operatorname{MHz}%
$, and $\zeta=6.0$ for the MW LA at $\omega/2\pi=3%
\operatorname{GHz}%
$. An impedance matching capacitor, serially connected to the RF solenoid, is
employed for the measurements shown in Fig. \ref{FigLZd}.

\begin{figure}[ptb]
\centering  \includegraphics[width=3.2in]{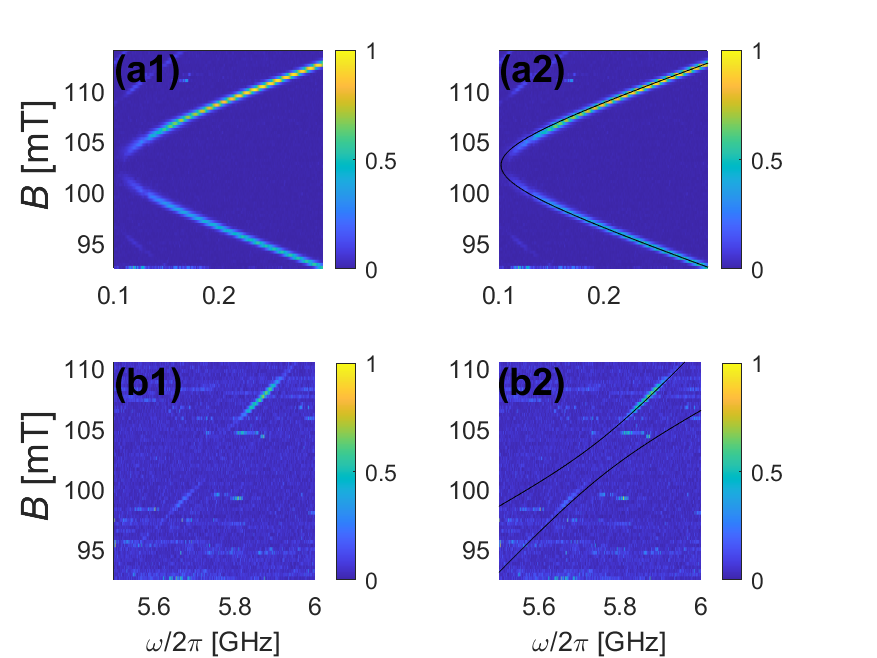} \caption{ODMR in the
GSLAC region. The low and high frequency resonances are shown in (a) and (b),
respectively. The black solid lines in (a2) and (b2) are based on numerical
diagonalization of the triplet Hamiltonian (\ref{H_NV TGS}).}%
\label{fig_ODMR}%
\end{figure}

The experimental setup used for performing ODMR measurements is described in
Ref. \cite{Alfasi_214111}. ODMR signal recorded near the GSLAC region is shown
in Fig. \ref{fig_ODMR} as a function of the driving frequency $\omega$ and
magnetic field $B$.

\begin{figure}[ptb]
\centering  \includegraphics[width=3.2in,keepaspectratio]{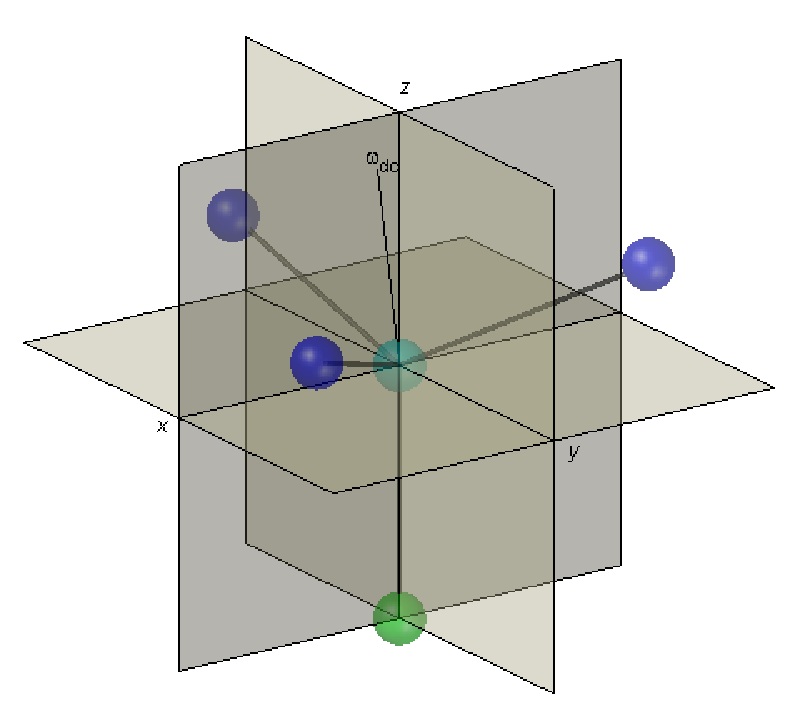}\caption{NV.
The NV axis is parallel to the $z$ axis, and the static field
$\boldsymbol{\omega}_{\mathrm{dc}}$, which is nearly parallel to the $z$ axis,
lies in the $xz$ plane.}%
\label{FigG}%
\end{figure}

\textbf{Triplet ground state} - When hyperfine interaction is disregarded the
ground state spin triplet Hamiltonian $\mathcal{H}$ becomes
\cite{Ovartchaiyapong_1403_4173,MacQuarrie_227602}%
\begin{equation}
\frac{\mathcal{H}}{\hbar}=\frac{\omega_{\mathrm{D}}S_{z}^{2}}{\hbar^{2}}%
+\frac{\omega_{\mathrm{E}}\left(  S_{+}^{2}+S_{-}^{2}\right)  }{2\hbar^{2}%
}-\frac{\gamma_{\mathrm{e}}\mathbf{B}\cdot\mathbf{S}}{\hbar}\;,
\label{H_NV TGS}%
\end{equation}
where $\omega_{\mathrm{D}}=2\pi\times2.87%
\operatorname{GHz}%
$ is the zero field splitting, $\omega_{\mathrm{E}}\ll\omega_{\mathrm{D}}$ is
a strain-induced splitting, $\gamma_{\mathrm{e}}=2\pi\times28.03%
\operatorname{GHz}%
\operatorname{T}%
^{-1}$ is the electron spin gyromagnetic ratio, $\mathbf{S}=S_{x}%
\mathbf{\hat{x}}+S_{y}\mathbf{\hat{y}}+S_{z}\mathbf{\hat{z}}$ is the spin
$S=1$ angular momentum vector operator and $S_{\pm}=S_{x}\pm iS_{y}$. The
applied magnetic field is expressed as $\mathbf{B}=\gamma_{\mathrm{e}}%
^{-1}\boldsymbol{\omega}$, where the angular frequency vector
$\boldsymbol{\omega
}=\left(  \omega_{x},\omega_{y},\omega_{z}\right)  =\boldsymbol{\omega
}_{\mathrm{dc}}+\boldsymbol{\omega}_{\mathrm{ac}}$ is decomposed into a static
part $\boldsymbol{\omega}_{\mathrm{dc}}=\left(  \omega_{\mathrm{dc},x}%
,\omega_{\mathrm{dc},y},\omega_{\mathrm{dc},z}\right)  $ (see Fig. \ref{FigG})
and a time varying part $\boldsymbol{\omega}_{\mathrm{ac}}=\left(
\omega_{\mathrm{ac},x},\omega_{\mathrm{ac},y},\omega_{\mathrm{ac},z}\right)  $
(having a vanishing averaged value). GSLAC occurs when
$\boldsymbol{\omega}\simeq\left(  0,0,\omega_{\mathrm{D}}\right)  $. The black
solid lines in Fig. \ref{fig_ODMR} (a2) and (b2) are derived from a numerical
diagonalization of the triplet Hamiltonian (\ref{H_NV TGS}).

\textbf{RWA} - The system's respond to both longitudinal and transverse
external driving is estimated below using the RWA. The $3\times3$\ matrix
representation $H=H_{\mathrm{dc}}+H_{\mathrm{ac}}$ of the Hamiltonian operator
$\mathcal{H}$ is decomposed into a static $H_{\mathrm{dc}}$ and time varying
$H_{\mathrm{ac}}$ parts. Two unitary transformations are successively applied
to $H$. The first one is performed with a time independent unitary matrix $U$,
which diagonalizes the static part, i.e. the $3\times3$ matrix $U^{\dag
}H_{\mathrm{dc}}U\equiv H_{\mathrm{dc}}^{\prime}$ is diagonal. The matrix
elements of $\hbar^{-1}H^{\prime}=\hbar^{-1}U^{-1}HU$ are denoted by
$\omega_{n^{\prime},n^{\prime\prime}}^{\prime}$, where $n^{\prime}%
,n^{\prime\prime}\in\left\{  1,2,3\right\}  $. The first transformation is
explained below in the next section (titled 'diagonalization') and in appendix
\ref{AppU}.

The second transformation, which is discussed in this section, and which is
applied to $H^{\prime}$, may help simplifying the evaluation of the steady
state response of the driven spin system under study. In general, nonlinear
response may give rise to a highly complex steady state. However, in some
cases the response is dominated by a single frequency mixing process. For such
cases the RWA can be employed to derive an analytical approximation for the
system's steady state response.

The second transformation is given by $H^{\prime\prime}=-i\hbar u^{\dag
}\left(  \mathrm{d}u/\mathrm{d}t\right)  +u^{\dag}H^{\prime}u^{{}}$, where the
elements $u_{n^{\prime},n^{\prime\prime}}$ of the diagonal unitary matrix $u$
are given by $u_{n^{\prime},n^{\prime\prime}}=\delta_{n^{\prime}%
,n^{\prime\prime}}\exp\left(  -i\int^{t}\mathrm{d}t^{\prime}\;\omega
_{n^{\prime},n^{\prime}}^{\prime}\left(  t^{\prime}\right)  \right)  $. The
matrix elements $\omega_{n^{\prime},n^{\prime\prime}}^{\prime\prime}$ of
$\hbar^{-1}H^{\prime\prime}$ are given by
\begin{equation}
\omega_{n^{\prime},n^{\prime\prime}}^{\prime\prime}=\left(  1-\delta
_{n^{\prime},n^{\prime\prime}}\right)  \omega_{n^{\prime},n^{\prime\prime}%
}^{\prime}e^{-i\int^{t}\mathrm{d}t^{\prime}\;\left(  \omega_{n^{\prime\prime
},n^{\prime\prime}}^{\prime}\left(  t^{\prime}\right)  -\omega_{n^{\prime
},n^{\prime}}^{\prime}\left(  t^{\prime}\right)  \right)  }\;.\label{omega''}%
\end{equation}
Note that the matrix $\hbar^{-1}H^{\prime\prime}$ is hollow, i.e. all its
diagonal elements $\omega_{n^{\prime},n^{\prime}}^{\prime\prime}$ vanish [see
Eq. (\ref{omega''})].

Consider the case where for some $n^{\prime}\neq n^{\prime\prime}$ the term
$\omega_{n^{\prime\prime},n^{\prime\prime}}^{\prime}-\omega_{n^{\prime
},n^{\prime}}^{\prime}$ in Eq. (\ref{omega''}) can be expressed as
$\omega_{n^{\prime\prime},n^{\prime\prime}}^{\prime}-\omega_{n^{\prime
},n^{\prime}}^{\prime}=\Omega_{0}-\Omega_{\mathrm{L}1}\cos\left(
\Omega_{\mathrm{L}}t^{\prime}\right)  $, and the term $\omega_{n^{\prime
},n^{\prime\prime}}^{\prime}$ as $\omega_{n^{\prime},n^{\prime\prime}}%
^{\prime}=\Omega_{\mathrm{T}1}e^{i\Omega_{\mathrm{T}}t}$, where $\Omega_{0}$,
$\Omega_{\mathrm{L}1}$, $\Omega_{\mathrm{L}}$, $\Omega_{\mathrm{T}1}$ and
$\Omega_{\mathrm{T}}$ are all real constants. With the help of the
Jacobi-Anger expansion one finds that for this case [see Eq. (\ref{omega''})]%
\begin{equation}
\omega_{n^{\prime},n^{\prime\prime}}^{\prime\prime}=\sum\limits_{l=-\infty
}^{\infty}\frac{\Omega_{1,l}}{2}e^{-i\Omega_{\mathrm{d},l}t}\;,
\label{omega'' Jacobi-Anger}%
\end{equation}
where $\Omega_{1,l}=2\Omega_{\mathrm{T}1}J_{l}\left(  \Omega_{\mathrm{L}%
1}/\Omega_{\mathrm{L}}\right)  $, $J_{l}$ denotes the Bessel function of the
first kind, and $\Omega_{\mathrm{d},l}$, which is given by $\Omega
_{\mathrm{d},l}=\Omega_{0}-\Omega_{\mathrm{T}}-l\Omega_{\mathrm{L}}$,
represents the angular frequency detuning corresponding to the $l$'th
frequency mixing resonance, occurring when $\Omega_{\mathrm{T}}+l\Omega
_{\mathrm{L}}=\Omega_{0}$.

Rapidly oscillating terms are disregarded in the RWA. Consider the case where
the matrix elements $\omega_{n^{\prime},n^{\prime\prime}}^{\prime\prime}$ of
$\hbar^{-1}H^{\prime\prime}$ are Fourier expanded as sums of terms having the
form $\left(  \Omega_{1}/2\right)  e^{-i\Omega_{\mathrm{d}}t}$ [see Eq.
(\ref{omega'' Jacobi-Anger})], where the driving amplitudes $\Omega_{1}$, and
the frequency detunings $\Omega_{\mathrm{d}}$, are real constants. The effect
of each such a term is strong when the amplitude $\left\vert \Omega
_{1}\right\vert $ is large, and the detuning $\left\vert \Omega_{\mathrm{d}%
}\right\vert $ is small. The analysis is greatly simplified when, among all
terms $\omega_{n^{\prime},n^{\prime\prime}}^{\prime\prime}$ in the upper
diagonal of $\hbar^{-1}H^{\prime\prime}$ (i.e. $n^{\prime}<n^{\prime\prime}$),
a single term having the form $\left(  \Omega_{1}/2\right)  e^{-i\Omega
_{\mathrm{d}}t}$ in a single matrix element $\omega_{n_{1},n_{2}}^{\prime
\prime}$, where $n_{1}<n_{2}$, dominates (recall that the diagonal elements of
$H^{\prime\prime}$ vanish). For this case, in the RWA all other terms are
disregarded, i.e. it is assumed that the only non-vanishing matrix elements
$\omega_{n^{\prime},n^{\prime\prime}}^{\prime\prime}$ of $\hbar^{-1}%
H^{\prime\prime}$\ are $\omega_{n_{1},n_{2}}^{\prime\prime}=\left(  \Omega
_{1}/2\right)  e^{-i\Omega_{\mathrm{d}}t}$ and $\omega_{n_{2},n_{1}}%
^{\prime\prime}=\left(  \Omega_{1}/2\right)  e^{i\Omega_{\mathrm{d}}t}$. The
effect of driving can be characterized by the coefficient $\mathcal{P}%
_{n_{1},n_{2}}$, which is given by [see Eq. (5) of Ref. \cite{Berns_150502},
and Eq. (D11) of Ref. \cite{Buks_033807}]%
\begin{equation}
\mathcal{P}_{n_{1},n_{2}}=\frac{\frac{\Omega_{1}^{2}}{\gamma_{1}\gamma_{2}}%
}{1+\frac{\Omega_{\mathrm{d}}^{2}}{\gamma_{2}^{2}}+\frac{\Omega_{1}^{2}%
}{\gamma_{1}\gamma_{2}}}\;, \label{P_n1,n2}%
\end{equation}
where $\gamma_{1}$ ($\gamma_{2}$) is a longitudinal (transverse) spin
relaxation rate. Note that for a two-level system $\mathcal{P}=1-P_{0}%
/P_{\mathrm{s}}$, where $P_{0}$ ($P_{\mathrm{s}}$) is the population
polarization in steady state with (without) driving [see
\cite{Slichter_Principles}, and Eq. (17.265) of Ref. \cite{Buks_QMLN}]. The
coefficient $\mathcal{P}_{n_{1},n_{2}}$ can be used to approximately quantify
the effect of driving on steady state ODMR signal.

\textbf{Diagonalization} - Prior to the above-explained transformation that
treats the driving terms using the RWA, a time-independent transformation is
applied in order to diagonalize the static part of the Hamiltonian
$H_{\mathrm{dc}}$. The diagonalized static part of the Hamiltonian reveals the
transition frequencies, whereas the unitary matrix that diagonalizes
$H_{\mathrm{dc}}$ reveals the transition selection rules.

The matrix $H$ is given by [the term $\omega_{\mathrm{zd}}/2$ has been added
to the diagonal, see Eq. (\ref{H_NV TGS})]%
\begin{equation}
\frac{H}{\hbar}=\left(
\begin{array}
[c]{ccc}%
-\frac{\omega_{\mathrm{zd}}}{2} & \frac{\omega_{\delta}^{{}}}{2} &
\omega_{\mathrm{E}}\\
\frac{\omega_{\delta}^{\ast}}{2} & \frac{\omega_{\mathrm{zd}}}{2} &
\frac{\omega_{\delta}^{{}}}{2}\\
\omega_{\mathrm{E}} & \frac{\omega_{\delta}^{\ast}}{2} & \omega_{\mathrm{h}}%
\end{array}
\right)  \;, \label{H / hbar}%
\end{equation}
where $\omega_{\mathrm{zd}}=\omega_{z}-\omega_{\mathrm{D}}$ represents
detuning from GSLAC, $\omega_{\delta}=-\sqrt{2}\left(  \omega_{x}-i\omega
_{y}\right)  $ represents transverse magnetic fields, and $\omega_{\mathrm{h}%
}=\left(  \omega_{\mathrm{D}}+3\omega_{\mathrm{zd}}\right)  /2$.

The static part $H_{\mathrm{dc}}$ of $H$ [see Eq. (\ref{H / hbar})] can be
expressed as (it is assumed that $\omega_{\mathrm{dc},y}=0$)%
\begin{equation}
\frac{H_{\mathrm{dc}}}{\hbar}=\frac{\omega_{\mathrm{R0}}}{2}\left(
\begin{array}
[c]{ccc}%
-\cos\theta & -\sin\theta & \frac{2\omega_{\mathrm{E}}}{\omega_{\mathrm{R0}}%
}\\
-\sin\theta & \cos\theta & -\frac{\omega_{\mathrm{H}}}{\omega_{\mathrm{R0}}}\\
\frac{2\omega_{\mathrm{E}}}{\omega_{\mathrm{R0}}} & -\frac{\omega_{\mathrm{H}%
}}{\omega_{\mathrm{R0}}} & \frac{2\omega_{\mathrm{dc,h}}}{\omega_{\mathrm{R0}%
}}%
\end{array}
\right)  \;, \label{H_dc}%
\end{equation}
where $\omega_{\mathrm{R0}}=\sqrt{\left(  \omega_{\mathrm{dc},z}%
-\omega_{\mathrm{D}}\right)  ^{2}+\omega_{\mathrm{H}}^{2}}$, $\omega
_{\mathrm{H}}=\sqrt{2}\omega_{\mathrm{dc},x}$, $\omega_{\mathrm{dc,h}}=\left(
\omega_{\mathrm{D}}+3\omega_{\mathrm{dc},z}\right)  /2$, and the angle
$\theta$ is given by%
\begin{equation}
\cot\theta=\frac{\omega_{\mathrm{dc},z}-\omega_{\mathrm{D}}}{\omega
_{\mathrm{H}}}\equiv\eta\;. \label{eta}%
\end{equation}
Note that GSLAC corresponds to the case where $\omega_{\mathrm{dc},z}%
=\omega_{\mathrm{D}}$, $\eta=0$ and $\left\vert \theta\right\vert =\pi/2$,
whereas $\left\vert \eta\right\vert \rightarrow\infty$ and $\theta
\rightarrow0$ far from GSLAC. The GSLAC transition angular frequency is
approximately given by $\omega_{\mathrm{R0}}=\omega_{\mathrm{H}}\sqrt
{1+\eta^{2}}$.

In the vicinity of the GSLAC, the static part $H_{\mathrm{dc}}$ of $H$ is
approximately diagonalized using the transformation $H_{\mathrm{dc}}^{\prime
}=U^{-1}H_{\mathrm{dc}}U$, where the unitary matrix $U$ represents a rotation
about the $z$\ axis through an angle of $\theta/2$. Applying the same
transformation $U$ to the time dependent part $H_{\mathrm{ac}}$ yields two
terms $H_{\mathrm{ac}}^{\prime}=U^{-1}H_{\mathrm{ac}}U=H_{\mathrm{acT}%
}^{\prime}+H_{\mathrm{acL}}^{\prime}$. The first one $H_{\mathrm{acT}}%
^{\prime}$\ , which is given by (see appendix \ref{AppU})
\begin{equation}
\frac{H_{\mathrm{acT}}^{\prime}}{\hbar}=\left(
\begin{array}
[c]{ccc}%
-\omega_{\mathrm{TL}} & \omega_{\mathrm{TT}}^{{}} & -\frac{\omega_{\mathrm{T}%
}^{\ast}\sin\frac{\theta}{2}}{\sqrt{2}}\\
\omega_{\mathrm{TT}}^{\ast} & \omega_{\mathrm{TL}} & -\frac{\omega
_{\mathrm{T}}^{\ast}\cos\frac{\theta}{2}}{\sqrt{2}}\\
-\frac{\omega_{\mathrm{T}}^{{}}\sin\frac{\theta}{2}}{\sqrt{2}} & -\frac
{\omega_{\mathrm{T}}^{{}}\cos\frac{\theta}{2}}{\sqrt{2}} & 0
\end{array}
\right)  \;, \label{H_ac,T}%
\end{equation}
where $\omega_{\mathrm{TL}}=2^{-3/2}\left(  \omega_{\mathrm{T}}^{{}}%
+\omega_{\mathrm{T}}^{\ast}\right)  \sin\theta$, $\omega_{\mathrm{TT}%
}=2^{-1/2}\left(  \omega_{\mathrm{T}}^{{}}\sin^{2}\left(  \theta/2\right)
-\omega_{\mathrm{T}}^{\ast}\cos^{2}\left(  \theta/2\right)  \right)  $, and
where $\omega_{\mathrm{T}}=\omega_{\mathrm{ac},x}+i\omega_{\mathrm{ac},y}$,
originates from transverse driving, whereas longitudinal driving gives rise to
the second term $H_{\mathrm{acL}}^{\prime}$, which is given by (see appendix
\ref{AppU})%
\begin{equation}
\frac{H_{\mathrm{acL}}^{\prime}}{\hbar}=\frac{\omega_{\mathrm{ac},z}}%
{2}\left(
\begin{array}
[c]{ccc}%
-\cos\theta & \sin\theta & 0\\
\sin\theta & \cos\theta & 0\\
0 & 0 & 3
\end{array}
\right)  \;. \label{H_ac,L}%
\end{equation}
Note that $\cos\theta=\eta/\sqrt{\eta^{2}+1}$ ,$\sin\theta=1/\sqrt{\eta^{2}%
+1}$ and $-\omega_{\mathrm{TL}}^{2}-\omega_{\mathrm{TT}}^{{}}\omega
_{\mathrm{TT}}^{\ast}=-\left\vert \omega_{\mathrm{T}}\right\vert ^{2}/2$
(determinant of the $2\times2$ top upper left block of $\hbar^{-1}%
H_{\mathrm{acT}}^{\prime}$).

\begin{figure}[ptb]
\begin{center}
\includegraphics[width=3.2in,keepaspectratio]{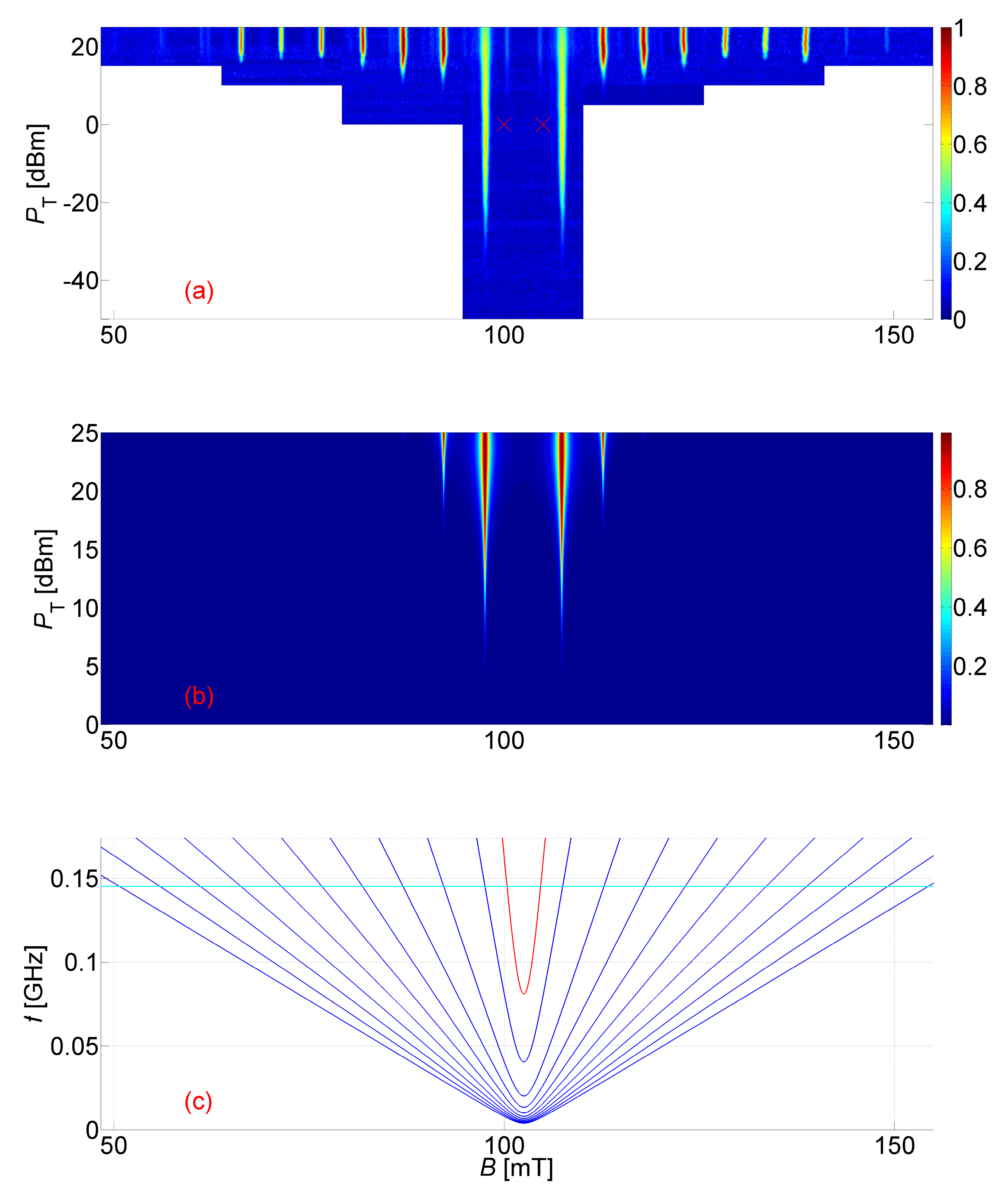}
\end{center}
\caption{Single antenna drive. (a) ODMR signal as a function of static
magnetic filed $B$ and driving power $P_{\mathrm{T}}$. (b) Calculated
driving-induced polarization coefficient $\mathcal{P}$ [see Eq. (\ref{P_n1,n2}%
)]. The parameters' assumed values for the calculation of $\mathcal{P}$ are
$\gamma_{1}=0.5 \operatorname{MHz} $ and $\gamma_{2}=2 \operatorname{MHz} $.
(c) Graphical representation of the frequency mixing resonance condition
$\omega_{\mathrm{R0}}=\omega_{\mathrm{H}}\sqrt{1+\eta^{2}}=l\omega
_{\mathrm{T}}$. The driving frequency $\omega_{\mathrm{T}}/\left(
2\pi\right)  =145 \operatorname{MHz} $ is represented by the horizontal cyan
line.}%
\label{FigLZ_Hyp}%
\end{figure}

\textbf{Single antenna drive} - The rotation angle $\theta/2$ associated with
the diagonalization of the static part $H_{\mathrm{dc}}$ (\ref{H_dc}) [see Eq.
(\ref{U=})] becomes relatively large near GSLAC. Consequently, in this region
the contribution of the MW LA to the effective value of longitudinal driving
amplitude becomes significant [see Eq. (\ref{H_ac,T})]. As is demonstrated
below (see Fig. \ref{FigLZ_Hyp}), this effective longitudinal driving gives
rise to a sequence of superharmonic resonances occurring near GSLAC.

The plot in Fig. \ref{FigLZ_Hyp}(a) exhibits ODMR measurements performed with
a monochromatic driving applied only to the MW LA. The driving has
a fixed frequency $\omega_{\mathrm{T}}/\left(  2\pi\right)  =145%
\operatorname{MHz}%
$ and a varying power $P_{\mathrm{T}}$. The driving amplitude, which is
denoted by $\omega_{\mathrm{T}1}$, is proportional to $P_{\mathrm{T}}^{1/2}$.
In the transformed basis (in which the static part of the Hamiltonian is
diagonalized), the effective longitudinal (transverse) driving amplitude is
$\omega_{\mathrm{TL}}$ ($\omega_{\mathrm{TT}}$) [see Eq. (\ref{H_ac,T})].

The $l$'th frequency mixing resonance occurs when $\omega_{\mathrm{R0}}%
=\omega_{\mathrm{H}}\sqrt{1+\eta^{2}}=l\omega_{\mathrm{T}}$, where $l$ is a
positive integer. This resonance condition is graphically displayed in Fig.
\ref{FigLZ_Hyp}(c). In the experimental data shown in Fig. \ref{FigLZ_Hyp}(a),
resonances with $l\leq10$ can be resolved. The driving-induced polarization
coefficient $\mathcal{P}$, which is calculated using Eqs.
(\ref{omega'' Jacobi-Anger}) and (\ref{P_n1,n2}), is shown in Fig.
\ref{FigLZ_Hyp}(b). Parameters' assumed values are listed in the caption of
Fig. \ref{FigLZ_Hyp}. The comparison between the measured ODMR signal (a) and
calculated polarization coefficient $\mathcal{P}$ (b) suggests that the decay
of resonance intensity with $\left\vert l\right\vert $ is theoretically
overestimated. The discrepancy is mainly attributed to dipolar coupling, which
gives rise to additional driving term, which is disregarded in the calculation
of $\mathcal{P}$ \cite{Anderson_1517,Cheng1961,Broer_801,Daycock1969}.

The two resonances labelled in Fig. \ref{FigLZ_Hyp}(a) by red cross symbols
represent second Larmor lines. The resonance condition for the second Larmor
lines, which is given by $2\omega_{\mathrm{R0}}=\omega_{\mathrm{T}}$, is
graphically shown in Fig. \ref{FigLZ_Hyp}(c) (the term $2\omega_{\mathrm{R0}}$
is represented by the red hyperbola, and $\omega_{\mathrm{T}}/\left(
2\pi\right)  =145%
\operatorname{MHz}%
$ by the horizontal cyan line). Note that the nonlinear process responsible
for the second Larmor lines is not taken into account in the calculation of
the driving-induced polarization coefficient $\mathcal{P}$ [see Fig.
\ref{FigLZ_Hyp}(b)].

\begin{figure}[ptb]
\begin{center}
\includegraphics[width=3.2in,keepaspectratio]{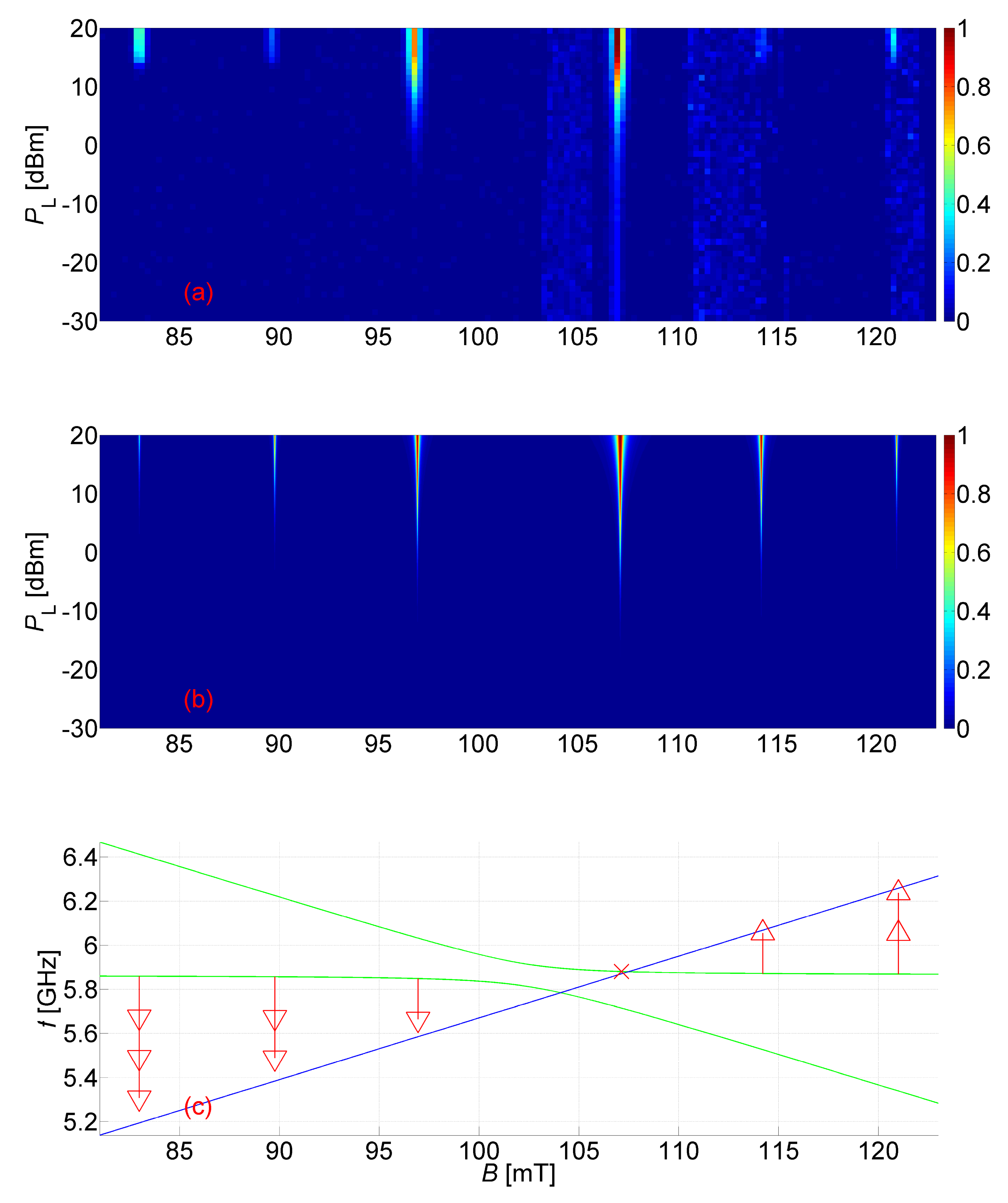}
\end{center}
\caption{Two antenna drive near GSLAC. (a) ODMR signal as a function of static
magnetic filed $B$ and RF solenoid driving power $P_{\mathrm{L}}$, with a
fixed RF solenoid driving frequency of $\omega_{\mathrm{L}}/\left(
2\pi\right)  =185 \operatorname{MHz} $, MW LA driving frequency of
$\omega_{\mathrm{T}}/\left(  2\pi\right)  =5.87 \operatorname{GHz} $ and MW LA
driving power of 25 dBm. (b) The calculated polarization coefficient
$\mathcal{P}$ [see Eq. (\ref{P_n1,n2})]. (c) Graphical representation of the
frequency mixing matching condition.}%
\label{FigLZ_T}%
\end{figure}

\textbf{Two antenna drive} - The case of simultaneous driving with both RF
solenoid and MW LA is exhibited by the plots shown in Figs. \ref{FigLZd} and
\ref{FigLZ_T}, which demonstrate that frequency mixing between off-resonance
monochromatic driving tones can be employed for resonance driving. This frequency mixing process is related to the Landau Zener St\"uckelberg
effect, which describes a coherent interference occurring in a two-level
system under simultaneously applied two driving tones \cite{Shevchenko_1}.
Landau Zener St\"uckelberg interferometry (see Fig. \ref{FigLZd}
below) commonly requires intense longitudinal driving. For exploring this
interferometry, a capacitor having capacitance of $36%
\operatorname{\mu F}%
$ is serially connected to the RF solenoid. The added capacitor helps
enhancing the magnetic field driving amplitude by suppressing the undesirable
effect of parasitic capacitance of the RF solenoid. The RF solenoid with the
added capacitor is operated near its resonance frequency of $10.5%
\operatorname{MHz}%
$.

For the ODMR signal shown in the plot in Fig. \ref{FigLZd}(a), the RF solenoid
is driven with a varying amplitude $\Omega_{\mathrm{L}1}$ (proportional to the
applied voltage amplitude $V_{\mathrm{RF}}$) and a fixed frequency of
$\Omega_{\mathrm{L}}/\left(  2\pi\right)  =10.5%
\operatorname{MHz}%
$, and the MW LA is driven with a fixed amplitude of $\Omega_{\mathrm{T}1}=5.6
\operatorname{MHz}$ and a fixed frequency of $\Omega_{\mathrm{T}}/\left(
2\pi\right)  =3.15%
\operatorname{GHz}%
$. The plot shown in Fig. \ref{FigLZd}(b) displays the calculated polarization
coefficient $\mathcal{P}$ [see Eq. (\ref{P_n1,n2})] in the same region, as a
function of $V_{\mathrm{RF}}$ and the current $I_{\mathrm{mag}}$ applied to
the solenoid generating the static magnetic field. The assumed parameters'
values used for the calculation of $\mathcal{P}$, which is base on Eqs.
(\ref{omega'' Jacobi-Anger}) and (\ref{P_n1,n2}), are listed in the caption of
Fig. \ref{FigLZd}.

While the measurements presented in Fig. \ref{FigLZd} have been performed away
from the GSLAC, the case of two antenna driving near GSLAC is demonstrated by
the plots shown in Fig. \ref{FigLZ_T}. The plot in Fig. \ref{FigLZ_T}(a)
displays ODMR data as a function of static magnetic filed $B$ and RF solenoid
driving power $P_{\mathrm{L}}$. For this measurement, the RF solenoid driving
frequency is $\omega_{\mathrm{L}}/\left(  2\pi\right)  =185%
\operatorname{MHz}%
$, MW LA driving frequency is $\omega_{\mathrm{T}}/\left(  2\pi\right)  =5.87%
\operatorname{GHz}%
$ and MW LA driving power is $25$ dBm. The plot shown in Fig. \ref{FigLZ_T}(b)
displays the calculated polarization coefficient $\mathcal{P}$ in the same
region [see Eqs. (\ref{omega'' Jacobi-Anger}) and (\ref{P_n1,n2})].

The calculated energy eigenvalues of the triplet Hamiltonian (\ref{H_NV TGS})
are plotted as a function of $B$ in Fig. \ref{FigLZ_T}(c). To graphically
display the frequency mixing matching condition, the two lowest eigen energies
[green solid lines in Fig. \ref{FigLZ_T}(c)] are vertically shifted upwards by
the MW LA driving frequency $\omega_{\mathrm{T}}/\left(  2\pi\right)  =5.87%
\operatorname{GHz}%
$. The highest eigen energy is represented by the solid blue line. The length
of the red vertical arrows in Fig. \ref{FigLZ_T}(c) is the RF solenoid driving
frequency $\omega_{\mathrm{L}}/\left(  2\pi\right)  =185%
\operatorname{MHz}%
$.

\textbf{Summary} - Nonlinear magnetic resonance sensing of NV$^{-}$ defects at
low temperatures is demonstrated using ODMR. Several frequency mixing
configurations are employed, including Landau Zener St\"uckelberg
interferometry and two-tone driving spectroscopy. Magnetic driving is applied
in the longitudinal and transverse directions using MW and RF fields. The
experimental results were compared with prediction of theoretical analysis
based on the RWA.

Frequency mixing offers some advantages for sensing applications, including
the ability to eliminate crosstalk between driving and detection. This
crosstalk problem is important for cavity-based detection of magnetic
resonance \cite{Alfasi_063808}. At cryogenic temperatures, the
method of frequency mixing can be used for applications such as magnetometery
of superconducting materials \cite{Alfasi_075311}. In this work,
ODMR has been employed for exploring the efficiency of a variety of frequency
mixing driving configurations. Future study will explore and optimize
protocols for cavity-based spin spectroscopy that are based on frequency mixing.

\textbf{Acknowledgments} - This work was supported by the Israeli science
foundation, and the Israeli ministry of science.

\appendix

\section{{Derivation of Eqs. (\ref{H_ac,T}) and (\ref{H_ac,L})}}

\label{AppU}

The approximate diagonalization of the static part $H_{\mathrm{dc}}$
(\ref{H_dc}) of $H$ in the vicinity of the GSLAC is based on the relation%
\begin{equation}
U^{-1}\left(
\begin{array}
[c]{ccc}%
-\cos\theta & -\sin\theta & 0\\
-\sin\theta & \cos\theta & 0\\
0 & 0 & 1
\end{array}
\right)  U=\left(
\begin{array}
[c]{ccc}%
-1 & 0 & 0\\
0 & 1 & 0\\
0 & 0 & 1
\end{array}
\right)  \;,
\end{equation}
where the unitary matrix $U$, which is given by%
\begin{equation}
U=\left(
\begin{array}
[c]{ccc}%
\cos\frac{\theta}{2} & -\sin\frac{\theta}{2} & 0\\
\sin\frac{\theta}{2} & \cos\frac{\theta}{2} & 0\\
0 & 0 & 1
\end{array}
\right)  \;, \label{U=}%
\end{equation}
represents a rotation about the $z$\ axis through an angle of $\theta/2$. The
derivation of {Eqs. (\ref{H_ac,T}) and (\ref{H_ac,L}) is performed using the
following matrix relations}%
\begin{equation}
U^{-1}\left(
\begin{array}
[c]{ccc}%
-1 & 0 & 0\\
0 & 1 & 0\\
0 & 0 & 0
\end{array}
\right)  U=\left(
\begin{array}
[c]{ccc}%
-\cos\theta & \sin\theta & 0\\
\sin\theta & \cos\theta & 0\\
0 & 0 & 0
\end{array}
\right)  \;,
\end{equation}%
\begin{equation}
U^{-1}\left(
\begin{array}
[c]{ccc}%
0 & 1 & 0\\
1 & 0 & 0\\
0 & 0 & 0
\end{array}
\right)  U=\left(
\begin{array}
[c]{ccc}%
\sin\theta & \cos\theta & 0\\
\cos\theta & -\sin\theta & 0\\
0 & 0 & 0
\end{array}
\right)  \;,
\end{equation}%
\begin{equation}
U^{-1}\left(
\begin{array}
[c]{ccc}%
0 & 0 & 1\\
0 & 0 & 0\\
1 & 0 & 0
\end{array}
\right)  U=\left(
\begin{array}
[c]{ccc}%
0 & 0 & \cos\frac{\theta}{2}\\
0 & 0 & -\sin\frac{\theta}{2}\\
\cos\frac{\theta}{2} & -\sin\frac{\theta}{2} & 0
\end{array}
\right)  \;,
\end{equation}%
\begin{equation}
U^{-1}\left(
\begin{array}
[c]{ccc}%
0 & 0 & 0\\
0 & 0 & 1\\
0 & 1 & 0
\end{array}
\right)  U=\left(
\begin{array}
[c]{ccc}%
0 & 0 & \sin\frac{\theta}{2}\\
0 & 0 & \cos\frac{\theta}{2}\\
\sin\frac{\theta}{2} & \cos\frac{\theta}{2} & 0
\end{array}
\right)  \;,
\end{equation}
and%
\begin{align}
&  U^{-1}\left(
\begin{array}
[c]{ccc}%
0 & x+iy & 0\\
x-iy & 0 & x+iy\\
0 & x-iy & 0
\end{array}
\right)  U\nonumber\\
&  =\left(
\begin{array}
[c]{ccc}%
x\sin\theta & x\cos\theta+iy & \left(  x+iy\right)  \sin\frac{\theta}{2}\\
x\cos\theta-iy & -x\sin\theta & \left(  x+iy\right)  \cos\frac{\theta}{2}\\
\left(  x-iy\right)  \sin\frac{\theta}{2} & \left(  x-iy\right)  \cos
\frac{\theta}{2} & 0
\end{array}
\right)  \;.\nonumber\\
&
\end{align}

\bibliographystyle{ieeepes}
\bibliography{acompat,Eyal_Bib}

\end{document}